\documentclass[twocolumn,superscriptaddress,showpacs,oneside,prl]{revtex4}%
\usepackage{amsmath}
\usepackage{amssymb}
\usepackage{amsfonts}
\usepackage{graphicx}%
\begin{document}
\title{Comment on ``Quantum Key Distribution with the Blind Polarization Bases''}
\author{Qiang Zhang}
\affiliation{Hefei National Laboratory for Physical Sciences at
Microscale \& Department of Modern Physics, University of Science
and Technology of China, Hefei, Anhui 230026, P.R. China}
\affiliation{Physikalisches Institut, Universit\"{a}t Heidelberg,
Philosophenweg 12, 69120 Heidelberg, Germany}
\author{Xiang-Bin Wang}
\affiliation{ Imai Quantum Computation and Information Project,
ERATO, Japan Science and Technology Agency, Daini Hongo White
Building 201, 5-28-3, Hongo, Bunkyo, Tokyo 113-033, Japan}
\author{Yu-Ao Chen}
\affiliation{Physikalisches Institut, Universit\"{a}t Heidelberg,
Philosophenweg 12, 69120 Heidelberg, Germany}
\author{Tao Yang}
\affiliation{Hefei National Laboratory for Physical Sciences at
Microscale \& Department of Modern Physics, University of Science
and Technology of China, Hefei, Anhui 230026, P.R. China}
\author{Jian-Wei Pan}
\affiliation{Hefei National Laboratory for Physical Sciences at
Microscale \& Department of Modern Physics, University of Science
and Technology of China, Hefei, Anhui 230026, P.R. China}
\affiliation{Physikalisches Institut, Universit\"{a}t Heidelberg,
Philosophenweg 12, 69120 Heidelberg, Germany} \maketitle

 In recent paper \cite{Kye}, Kye \emph{et al.} claim that using the blind polarization,
their new quantum key distribution scheme can be secure even when
a key is embedded in a not-so-weak coherent-state pulse. Here we
show an eavesdropping scheme, by which the eavesdropper can
achieve the full information of the key with a probability of
unity and will not be discovered by the the legitimate users, even
in the case that they have the perfect single-photon source and
the loseless channel.

There are two protocols in Ref. \cite{Kye}. Consider their first
protocol:
\\
{\bf Protocol 1:}

($a1$) Alice(the sender) prepares a linear polarized qubit in its
initial state $|\psi\rangle_{0} = |0\rangle$, where $|0\rangle$
and $|1\rangle$ represent two orthogonal polarizations of the
qubit, and chooses a random angle $\theta$.

($a2$) Alice rotates the polarization of the qubit by $\theta$, to
bring the state of the qubit to $|\psi\rangle_{1}
=\hat{U}_{y}(\theta)|\psi\rangle_{0} = \cos\theta|0\rangle
-\sin\theta|1\rangle$, and then sends the qubit to Bob(the
receiver).

($a3$) Bob chooses another random angle  and rotates the
polarization of the received qubit by $\phi$; $|\psi\rangle_{2}
=\hat{U}_{y}(\phi)|\psi\rangle_{1} = \cos(\theta+\phi)|0\rangle
-\sin(\theta+\phi)|1\rangle$, Bob sends the qubit back to Alice.

($a4$) Alice rotates the polarization angle of the qubit by
$-\theta$ and then encodes the message by further rotating the
polarization angle of $\pm\frac{\pi}{4}$; $|\psi\rangle_{3}
=\hat{U}_{y}(\pm\frac{\pi}{4})\times\hat{U}_{y}(-\theta)|\psi\rangle_{2}
$, Alice send the qubit to Bob. (Alice and Bob have predetermined
that $+\frac{\pi}{4}$ is say, ``0" and
 $-\frac{\pi}{4}$ is ``1".

($a5$) Bob measure the polarization after rotating the
polarization by $-\phi$; $|\psi\rangle_{4}
=\hat{U}_{y}(-\phi)|\psi\rangle_{3}=\hat{U}_{y}(\pm\frac{\pi}{4})|\psi\rangle_{0}
$, $\hat{U}_{y}(+\frac{\pi}{4})$ and $\hat{U}_{y}(-\frac{\pi}{4})$
are orthogonal to each other, which enables Bob to read the keys
precisely.

Our attacking scheme works as following:

($a1'$)  After step ($a2$) in the above protocol, Eve intercepts
all qubits from Alice and stores them. (We denote these qubits as
``set 1").
 Meanwhile, sends her own qubits with each of them being randomly in state
$|p\rangle$, $p$ is either 0 or 1. (We denote these qubits as
``set 2"). Eve can remember the state of each qubits sent from
her.

($a2'$) After step ($a3$), Eve intercepts all qubits from Bob and
stores them. Note that all these qubits are from set 2 originally.
Then, Eve sends set 1 to Alice.

($a3'$) After step ($a4$), Eve intercepts all qubits from Alice
and measures each of them in $\pm \frac{\pi}{4}$ basis.
 Reading the measurement outcome, Eve knows Alice's
choice of $\frac{\pi}{4}$ or $-\frac{\pi}{4}$, i.e., the bit
values of each pulse from Alice, $k=0,1$ for $\frac{\pi}{4}$ or
$-\frac{\pi}{4}$, respectively.

($a4'$) According to the measurement result of each qubits from
Alice, Eve takes appropriate unitary rotations to those qubits
stored by her in step ($a2'$) and sends them to Bob. Explicitly,
to any qubit, if its original value $p=0$, Eve rotates the
polarization by $(-1)^k\frac{\pi}{4}$ and sends it to Bob; if
$p=1$, she first flips its polarization between $|0\rangle$ and
$|1\rangle$ and then rotates the polarization by
$(-1)^k\frac{\pi}{4}$ and sends it to Bob. Bob will just implement
step ($a5$) and in such a way, Bob will have no error after the
protocol.

Knowing that protocol 1 suffers from the above types of
impersonation attack \cite{Kye}, they have proposed protocol 2 and
claimed that the modified protocol will be secure under such
attacks. We now show that Eve can access the full information of
the key from their modified protocol by almost the same
impersonation attack. Let us first recall their modified protocol.
\\
{\bf Protocol 2:}

($b1$) Alice sends two single-photon pulses of the polarization
angles $\theta_{1}$ and $\theta_{2}$ to Bob.

(b2) Bob rotates the polarization angles of the pulses by
$\phi+(-1)^{s}\frac{\pi}{4}$ and
$\phi+(-1)^{s\oplus1}\frac{\pi}{4}$ ,where the shuffling parameter
$s\in\{0,1\}$ is randonmly chosen by Bob and $\oplus$ denotes
addition modulo 2. And then Bob sends the pulses back to Alice.

(b3) Receiving the two pulses of their polarization angles
$\theta_{1}+\phi+(-1)^{s}\frac{\pi}{4}$ and
$\theta_{2}+\phi+(-1)^{s\oplus1}\frac{\pi}{4}$, Alice rotates the
polarization angles of the pulses by
 $-\theta_{1}+(-1)^{k}\frac{\pi}{4}$ and
 $-\theta_{2}+(-1)^{k}\frac{\pi}{4}$ respectively, where
 $k\in\{0,1\}$ is the key value. She blocks one of the qubits and
 sends the other to Bob. The paper \cite{Kye} introduces the blocking
 factor $b=0,1$ to denote the case to let the first or the second pulse  go.

 ($b4$) When the qubit travels to Bob, he rotates the polarization angles of the
pulses by
 $-\phi$ and measures the polarization. He obtains the measurement
 outcome $l^{sk}=s\oplus k\oplus b$ as the prekey bit value.

 ($b5$) Alice publicly announces her blocking factor b. And
 depending on b and the shuffling parameter s, Bob can decode the
 original key bit by $k=s\oplus b\oplus l$.

 Our attacking scheme is now the following:

 ($b1'$) After step ($b1$), Eve intercepts and stores both  pulses
from Alice. (We denote these pulses as ``set $E1$''). Eve also
sends two pulses to Bob. These two pulses are originally produced
by Eve with random polarization angles of ${\theta_{1}}'$,
${\theta_2}'$. Eve remembers the polarization angle of each
pulses.

 ($b2'$) After step ($b2$) in protocol 2, Eve intercepts both pulses
 from Bob, rotates each of them by angle $-{\theta_{1}}',-{\theta_2}' $
 and stores them. (We denote these pulses as ``set $E2$'').
Eve chooses her shuffling parameter $s'=0$ and rotates the two
pulses in set $E1$ by angle $(-1)^{s'}\frac{\pi}{4}$ and
$(-1)^{s'\bigoplus1}\frac{\pi}{4}$ respectively. Eve sends the
rotated pulses in set $E1$ to Alice.

 ($b3'$) After step ($b3$), Eve intercepts and measures the pulse sent out
by Alice in $|0, \frac{\pi}{2}\rangle$ basis.
 By reading measurement outcome, she knows the value
 $l^{s'k}=s'\oplus k\oplus b=k\oplus b$
since the value $s'$ is set by herself. (In our attacking scheme,
$s'=0$). Then she rotates the first
 pulse of set $E2$  with the angle of $(-1)^{k\oplus b}\frac{\pi}{4}$,
sends it to Bob and discards the second pulse in set $E2$.
(Remark: Although Eve knows neither $b$ nor $k$, she knows the
value of $k\oplus b$ and she can do the unitary rotation dependent
on $k\oplus b$).

When Alice announces the value of $b$ i.e., which pulse she has
blocked, Eve can get the value of k since she has already known
the value of $k\oplus b$ and $k=b\oplus (k\oplus b)$. On the other
hand, with the operation in step ($b3'$), Bob's result about
$k-$value will be identical to Alice's.
 Alice and Bob can not discover the eavesdropping. When Bob
 receives the pulse after step ($b3'$), he makes the measurement like
 step ($b5$) and will get the value  $l^{sk}=s\oplus (k\oplus b)$.
Although the $b'$ value chosen by Eve may be different from the
$b$ value chosen by Alice, it does not cause any channel noise
since Eve's actions in our attacking scheme will not affect Bob's
measurement outcome for the value $l$, which is $l=s\oplus k\oplus
b$.

Although it is known protocol 1 is probably insecure under
impersonation attack \cite{Kye}, it has been assumed that protocol
2 is secure prior to our comment. We have for the first time shown
the Protocol is totally insecure under our attack. Our
eavesdropping method works even in the case that Alice has a
perfect single-photon source and a transparent, noiseless channel.
The protocols of blind polarization bases are insecure in their
present forms \cite{Kye}. Moreover, in our present scheme, the
quantum memories which are not practical in nowadays technology
are not necessary, a long enough delay of the photon will be
sufficient. Therefore, our attacking scheme only need existing
technology.

\end{document}